\documentclass{kluwer}    
 \usepackage[dvips]{graphicx}

 \newdisplay{guess}{Conjecture}

\begin{document}
 \begin{article}
 \begin{opening}
 \title{Scattering Properties and Composition of Cometary Dust}
 \author{Ranjan \surname{Gupta}\email{rag@iucaa.ernet.in}}
 \institute{IUCAA, Post Bag 4, Ganeshkhind, Pune-411007, India}
 \author{D.B. \surname{Vaidya}\email{dbv@satyam.net.in}}
 \institute{Gujarat College, Ahmedabad-380006, India}
 \author{J.S. \surname{Dobbie}}
 \institute{Inst. for Atmospheric Science, University of Leeds, Leeds LS2
 JT,
 UK}
 \author{Petr \surname{Chylek}}
 \institute{Space and Remote Sensing Sciences, Los Alamos National
 Laboratory, Los Alamos, NM 87545, USA}
 \runningauthor{Gupta, Vaidya, Dobbie \& Chylek}
 \runningtitle{Scattering, composition of cometary dust}
 \date{September 15, 2003}

 \begin{abstract}

 Composition of the Comet dust obtained by the dust impact analyzer
 on the Halley probes indicated that the comet dust is a mixture of
 silicate and carbonaceous material. The collected interplanetary
 dust particles (IDP's) are fluffy and composite, having grains of
 several different types stuck together. Using Discrete Dipole
 Approximation (DDA) we study the scattering properties of composite
 grains. In particular, we study the angular distribution of the
 scattered intensity and linear polarization of composite grains.
 We assume that the composite grains are made up of
 a host silicate sphere/spheroid with the inclusions of graphite.
 Results of our calculations on the composite grains  show that
 the angle of maximum polarization shifts, and the degree
 of polarization varies with the volume fraction
 of the inclusions.
 We use these results on the composite grains to interpret
the observed scattering in cometary dust.
  \end{abstract}

 \end{opening}

 \section{Introduction}

 {\it {Insitu}} sampling of comet dust composition obtained by
       the dust impact analyzer on the Halley probes indicated a
       composition which is a mixture of silicate and
  carbonaceous material
       (Kissel {\it {et al.}} 1986, Jessberger {\it {et al.}} 1988,
  Jessberger 1999).
  The collected
       interplanetary dust particles (IDPs) of likely cometary origin
        are fluffy and composite (Brownlee 1987) and are typically
  of submicron to micron sized silicate grains
        embedded in a carbon rich matrix
          (Bradley {\it {et al.}} 1992).
The laboratory studies of the composite particles (Chylek {\it {et al.}}
 1988) and aggregates (Gustafson and Kolokolova 1999)
        show that the scattering properties
        for these irregular and inhomogeneous particles are
        different from that of solid homogeneous particles. Mie theory
 for spherical particles is not capable of explaining the scattering
 properties
 of cometary dust particles because cometary particles are fluffy and
 nonspherical
 (see e.g. Jockers 1997).
 Hence, there is a need to formulate models of electromagnetic
scattering by the inhomogeneous and nonspherical grains.
Calculations based on the discrete dipole approximation
(DDA) (Purcell and Pennypacker 1973, Draine 1988, Lumme {\it {et
al.}}
 1997,
  Wolff {\it {et al.}} 1998)
       allow the study of the light scattering
       properties of the irregularly shaped and inhomogeneous grains.
      Xing and Hanner (1997) and Yanamandra-Fisher and Hanner (1999)
  have studied the
  scattering properties of the aggregates of silicates and carbonaceous
 material to
  interpret the observed polarization in comets. However, the shape of the
  polarization curve for the cometary dust is still not very well
explained
  and also it is not very clear how the
  silicate and absorbing grains are mixed in the cometary dust (Hanner
2002).
Petrova {\it {et al.}} (2000) have shown that aggregates composed of touching
spheres with size parameters 1.3-1.65 display properties typical
of cometary particles; viz. a weak increase of the back scattering
intensity, negative linear polarization at small phase 
angles ($\leq20^{\circ}$) and a positive
wavelength gradient polarization. Their results on the aggregates
indicate that more compact particles have a more pronounced negative
branch of polarization.
 Lamy {\it {et al.}} (1987) had suggested a composite grain model of
silicate
 and graphite 'rough' grains to fit the observed polarization curve in the
 dust coma of comet Halley. The polarization of comet Halley was also
 observed at near IR wavelengths (viz. 1.25, 1.65 and 2.2 $\mu m$) at
several
 phase angles by Brooke {\it {et al.}} (1987).  They found that P/Halley
 exhibited
 a negative linear polarization branch at small phase angles i.e.
$\leq20^{\circ}$.
 They modeled  their observations using solid Mie particles
 with two different compositions, one component being of very absorbing
 material. However they suggested a two component  grain model with
 irregular grains for better understanding of the scattering properties of
 cometary dust. Jones {\it {et al.}} (2000) have observed the polarization
 of dust in Comet Hale-Bopp in the K-band ($\rm 2.2\mu m$) and they
 found the negative branch of polarization
 was much reduced (almost negligible) in the K band compared to that
 was observed in the visual wavelength range. Kelley {\it {et al.}}(2004)
have observed several comets in the K-band (i.e. $\rm 2.2\mu m$)
and have found that the K-band polarization was only 1\%-2\% higher
than typical optical polarization observed for the comets
(Levasseur-Regourd {\it {et al.}} 1996). Kelley {\it {et al.}}(2004) 
also found that the comet
Hale-Bopp showed no significant negative polarization branch
at $\rm 2.2\mu m$ at large scattering angles.
For cometary dust the radiative energy received from Sun is either
absorbed/
and in part re-radiated in the thermal infrared wavelength range or is
scattered.
Harker {\it {et al.}} (2002) have used a fractal porous model of the amorphous
carbon
      and amorphous silicate grains to model thermal emission from the
dust
coma
      of comet Hale-Bopp. In order to produce the fractal porous grain
model, Harker {\it {et al.}} (2002)
      have used the effective medium approximation (EMA). However, it is
to
be
      noted that the EMA does not take into account the inhomogeneities
within
the structure or the surface roughness of the grains (see for example
Wolff {\it {et al.}} 1998).

      In the present study we are concerned with the scattered part of the
solar
      radiation and it can be observed in the uv, visual and near infrared
wavelength
      ranges upto about 3$\mu m$; when the thermal radiation starts to
become
      important (see for example Jockers 1997).

  In this paper we use the results of the DDA calculations on the composite grains with various mixing ratios of the
 constituent
  materials to explain the observed scattering properties,
namely the angular distribution of the scattered intensity
and linear polarization of cometary dust.

 DDA has been used earlier by us to study the scattering properties of a
  composite particle made up of a host water sphere with carbon inclusions
  (Chylek {\it et al.} 2000). Recently we have used DDA to
  study the extinction properties
   of the composite grains (Vaidya {\it {et al.}} 2001).  In the present
 study,
  we calculate the angular distribution of the scattered intensity and
linear
polarization for the spherical and non-spherical composite grains with
silicate and graphite as constituent materials. The composite grains are
assumed to
  be made of the host silicate sphere (or spheroid)
      with inclusions of graphite.
      We study the effect of the inclusion size
     and volume fraction on the angular distribution
     of the
     scattered intensity and the linear polarization for the composite
 grains.
 It is to be noted that the composite grain model we present here for
 cometary
dust is different from the aggregate grain model of monomers suggested by
 Xing and Hanner (1997).
  In section 2, we present the composite grain models. In section 3 we
 give
  the results of our calculations and compare the results of the models
ith
 the observations.
  Summary and conclusions are given in the last section 4.

 \section{Discrete Dipole Approximation (DDA) and composite grain model}

     The basic DDA method consists of replacing
    a particle by an array of N oscillating polarizable point dipoles
  (see e.g. Draine 1988).
   The dipoles are located on a lattice and polarizability is related to
the
   complex refractive index m through a lattice dispersion relationship
   (Draine and Goodman 1993). Each dipole responds to the external
incident
  electric field and also to the electric fields of the other N-1 dipoles
 that
  comprise the particle. The polarization at each dipole site is
therefore
  coupled to all other dipoles in the particle.  There are two validity
  criteria for the DDA to provide
  accurate results (see e.g. Draine and Flatau 1994, Wolff {\it {et al.}}
 1994);
   viz (i) $\rm |m| k d \leq 1$,
  where m is the complex refractive index, k is the wave number,
  $\rm k=\pi/\lambda$ and d is the distance between the dipoles;
  and (ii) d should be small enough (N should be sufficiently large)
  to describe the shape of the particle
  satisfactorily.
  In the present study we use the modified DDA code
  (Dobbie 1999, Vaidya {\it {et al.}} 2001) to
     calculate the angular distribution of the scattered intensity
   and the linear polarization P for a composite grain.
 The code, first carves out an outer sphere (or spheroid) from a lattice
of
 dipole sites. Sites outside the sphere are vacuum and sites inside are
 assigned
 to be the host material. Once the host particle is formed the code
locates
 centers for internal spheres to form the inclusions. The inclusions are
of
a
 single radius and their centers are chosen randomly. The code then
outputs
  a three dimensional matrix specifying the material
  type at each dipole site which is then read by the DDA.
  In the present case, the sites are either silicates, graphites or
vacuum.
 The porosity P of the composite grain is defined as the relative amount
 of volume filled by the vacuum in the composite grain and is given by
  $\rm p = 1 - V_{solid}/V_{total}$, where $\rm V_{solid}$ is
 the solid material inside the grain and $\rm V_{total}$ is the total
volume of
 the grain (Hage and Greenberg 1990a \& 1990b).
 Accordingly, the porosity p for the composite grains varies between
 $\rm 0 < p < 1 $ (for more details on porosity and porous grains
see Vaidya \& Gupta, 1997 \&1999).
  Using the modified code for the composite grains, we have studied
  composite grain models with a host sphere (or spheroid) containing
   N=57856, 14440 and 25896 dipoles each carved out from
   $48\times48\times48$ , $48\times24\times24$ and $48\times32\times32$
   dipole sites respectively. 
 The composite grain models with N=1184 and 152 (p = 60\% and 80\%
respectively) 
are also generated and are used to study the effect of porosity on 
the scattering properties of the composite
grains .
      The volume fractions of the graphite inclusions used
are
 10\%, 20\%
     and 30\%. The inclusion size is labeled by the
   number of dipoles n across the diameter of an inclusion (see Table 1)
 (Chylek {\it {et al.}} 2000).
  The equivalent volume
radius for the inclusion is given by $\rm r_{eq} = (3 n /4 \pi)^{1/3} d$.
  For more details on the computer code and the corresponding
modification
  in the DDA code, see Dobbie (1999).

  Figures 1a and 1b illustrate the composite grain model for a
  typical spheroid (N=14440), used in the present study.
  Figure 1a shows the host spheroid (sites are shown in green
  color circles) with the embedded inclusions (shown in red color
circles).
  Since all the embedded inclusions cannot be seen and only the ones
  at the outer periphery are visible (Figure 1a), in the Figure 1b we
show
 the
  positions of all these inclusions in the host grain.

 It is to be emphasized that in the composite grain model
 considered in the present study, the inclusions
 are
 embedded in the host sphere, their positions are chosen randomly and
they
 are not
 overlapping; whereas in the aggregate grain model (Xing and
 Hanner 1997) of monomers, one monomer
 in the
 center is surrounded by the rest of the monomers; which are either
 overlapping, touching
 or separated.

      The optical constants for silicate and graphite;
  ($\rm m = 1.71 + 0.033i$ and $\rm 4.04 + 2.58i$ respectively) at
 wavelength
  of $2.2\mu m$,
 used in the present calculations are obtained from Draine (1985).
  Table 1 shows the parameters for the composite grain models considered
  in the present study. Here Si+Gr denotes
  the composite grain having silicate as the material for the host sphere
or
 spheroid
  and graphite as the material for the inclusion.

 \begin{figure}[H]
 \tabcapfont
 \centerline{%
 \begin{tabular}{c@{\hspace{0pc}}c}
 \includegraphics[width=15pc]{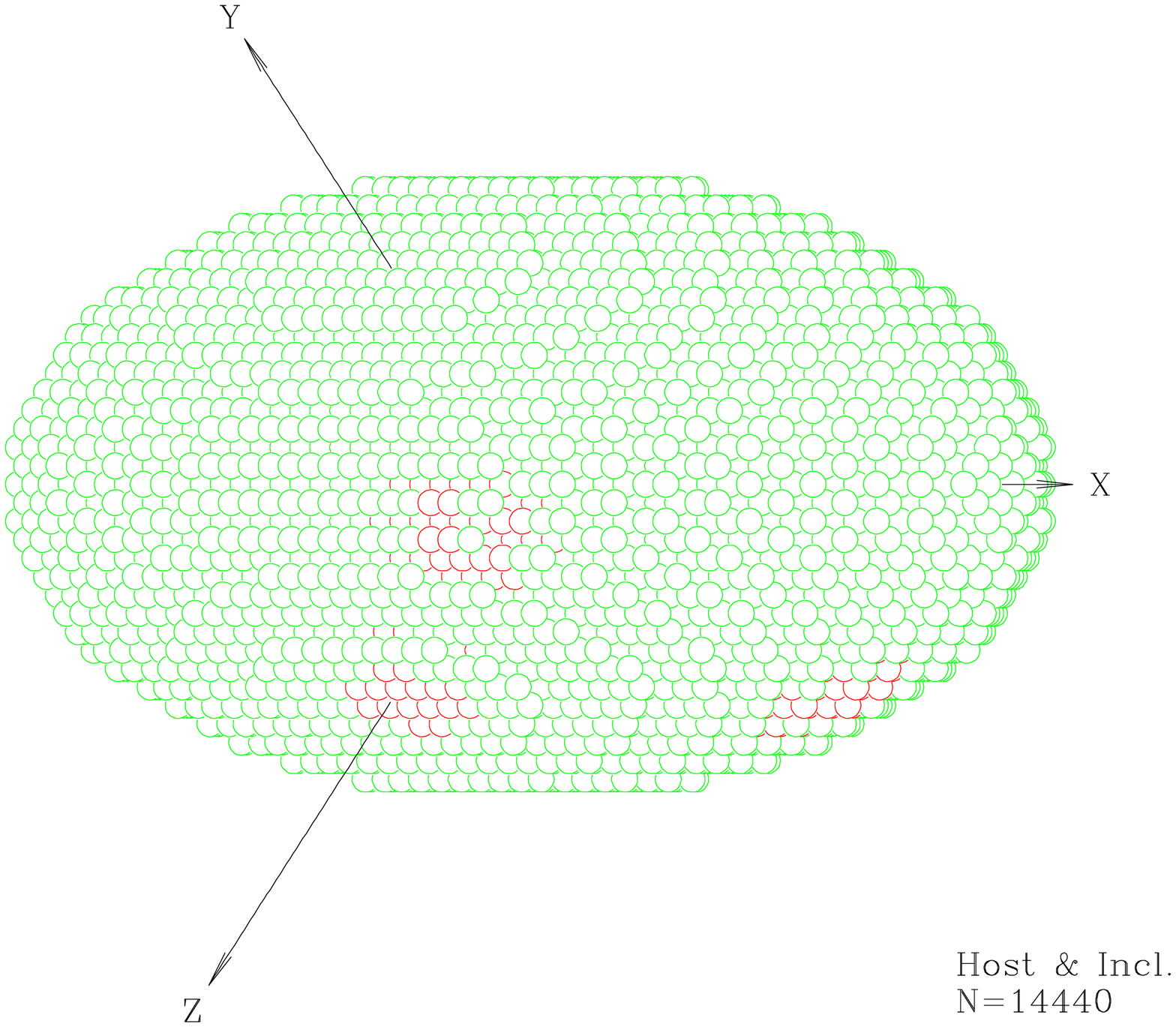} &
 \includegraphics[width=15pc]{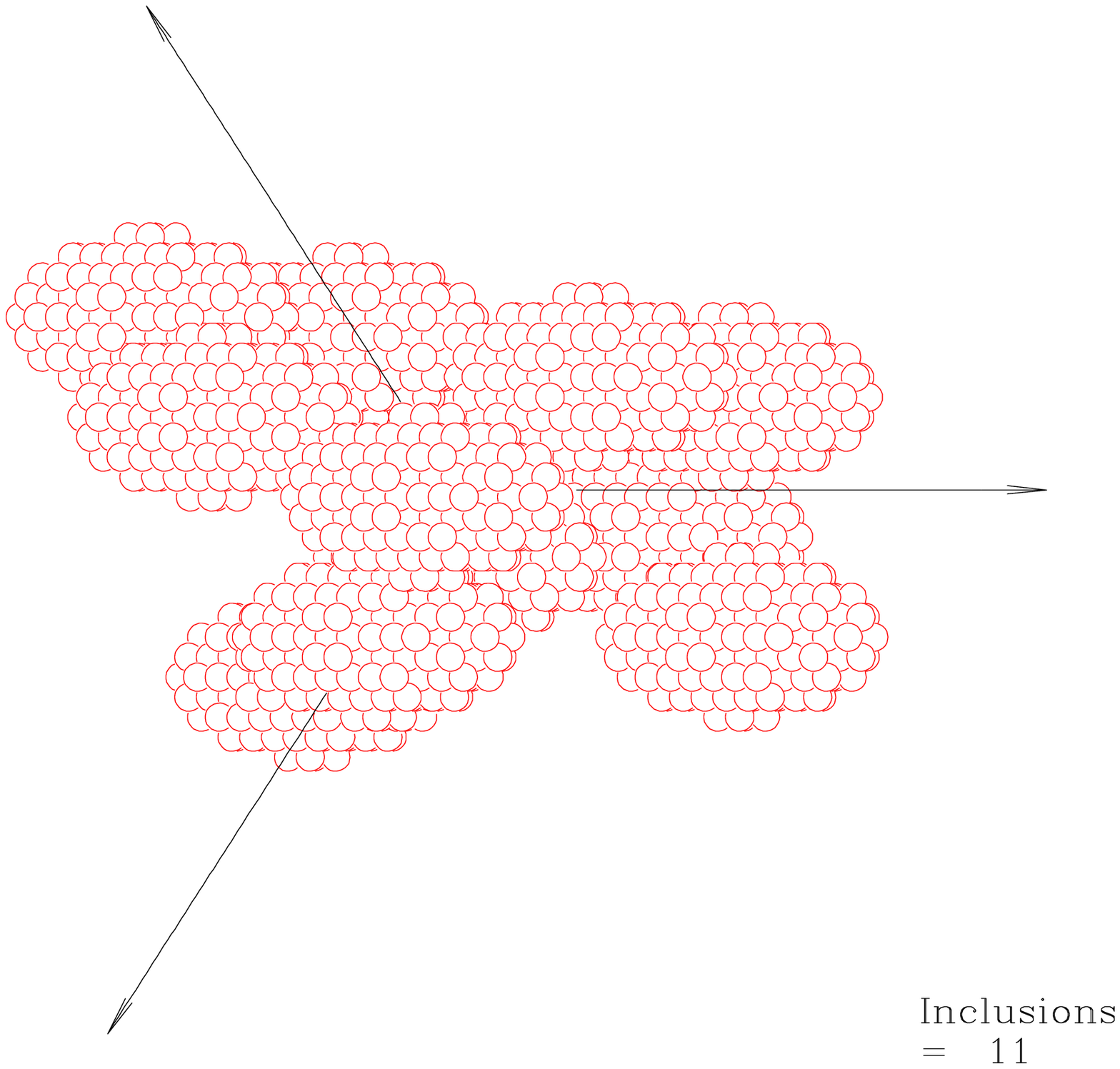} \\
 a. & b.
 \end{tabular}}

 \caption{A typical Non-spherical Composite grain
  with a total of N=14440 dipoles. (a) shows the inclusions embedded
  in the host spheroid such that only the ones placed at outer periphery
 are seen and (b) shows the inclusions.}
 \end{figure}

  \begin{table*}
  \begin{center}
  \caption{\bf{Composite Grain Models}}
  \vspace{1.0cm}
  \begin{tabular}{|c|c|c|c|} \hline
  Shape of & Number of & Volume Fraction & Size of \\
  Host Grain & Dipoles & of Inclusions & Inclusions \\
   & N & f & n  \\ \hline

     Sphere &      57856 &             0.1 &        12 \\
     (Si+Gr)&            &            0.2 &       12\\
            &            &            0.3 &       12\\

            &            &            0.2 &       6\\
            &            &            0.2 &       8\\ \hline

     Spheroids &   14440 &            0.1 &       12\\
     (Si+Gr)   &         &            0.2 &       12\\
               &         &            0.3 &       12\\

               &         &            0.2 &       6\\
               &         &            0.2 &       8\\  \hline

    Spheroids  &   25896 &            0.1 &       12\\
    (Si+Gr)    &         &            0.2 &       12\\
               &         &            0.3 &       12\\ \hline

    Spheroids  &   1184  &            0.2 &       6\\
    (Si+Gr)    &         &            &       \\ \hline

    Spheroids  &    152  &            0.2 &       4\\
    (Si+Gr)    &         &                &    \\ \hline
  \end{tabular}
  \end{center}
  \end{table*}

    We have used the size range $\rm a_{eq}$
   from  0.05 to 1.0$\mu$ , which corresponds
    to equivalent volume size parameter $\rm X=2\pi a_{eq}/\lambda$ from
  0.14 to 3.0 at the wavelength
    of $2.2\mu m$ , where $\rm a_{eq}$ is
 the radius of the
  sphere of equivalent volume of the host grain. For all the grain models
and
 grain sizes considered in the present study, the DDA criterion
 $\rm |m| k d \leq 1$
  is satisfied.

   To model the randomly oriented grains it is necessary to get the
  scattering properties
   of the composite grains averaged over all of the possible
orientations;
  i.e.

 $\rm \beta(0 \rightarrow 360^{\circ})$,  $\rm \theta(0 \rightarrow
180^{\circ})$ and  $\rm \phi(0 \rightarrow 360^{\circ})$;
 (see Draine and Flatau 1998).
   In the present
   study we use three values for each of the orientation
  parameters ($\rm \beta,  \theta,  \phi$)
   i.e. averaging over 27 orientations; which we find quite adequate;
higher
 number
 of orientations (e.g. Xing and Hanner 1997, Yanamandra-Fisher and Hanner
 1999)
 would require considerable computer cpu time.
It should be noted that we have also used  grain size distributions for
averaging
and folding
up the curves at $0^{\circ}$ and $180^{\circ}$ scattering angle
(see for example Jockers 1997).

 \section{Results \& Discussion}

    To be consistent with the comet dust composition, i.e. silicate and
   absorbing
     material with high carbon content (Kissel {\it {et al.}} 1986),
     we have selected silicate and graphite as the constituent materials
         for the composite grains in the present study.
 Earlier, Vanysek and Wickramasinghe (1975) had also suggested
 silicate and graphite particles as possible
 constituents in the cometary material.
 A two component grain model with
 'rough' grains was also suggested by Lamy {\it {et al.}} (1987)
 and Brooke {\it {et al.}} (1987)
 to fit the observed polarization in the dust coma of comet Halley.
  In this paper we have made a systematic study of
  the effect of variation in the volume fraction and the size of the
  inclusions on the scattering properties of the composite
  grains.
  The composite grain models considered in the present study are
  shown in Table 1.

  We have considered spherical and non-spherical composite grains in the
size
  parameter range X from 0.14 to 3.0, corresponding to the radius $\rm
a_{eq}$
 of
  0.05 to 1.0$\mu$ of the host grain at the wavelength of 2.2$\mu m$.

  {\it{Spheres:}}

   Figures 2(a-f) show the angular scattered intensity (i.e. S11) and the
    linear polarization P (i.e. --S12/S11)
   (Bohren and Huffman 1983)
   for the composite grains with the host silicate sphere,
   containing N= 57856 dipoles and for three
   volume fractions (viz. 10\%, 20\% and 30\%) of graphite inclusions
   for a constant inclusion size (n=12, i.e. $\rm r_{eq} \sim 1.45 d$).

  \begin{figure}
 \centerline{\includegraphics[width=25pc]{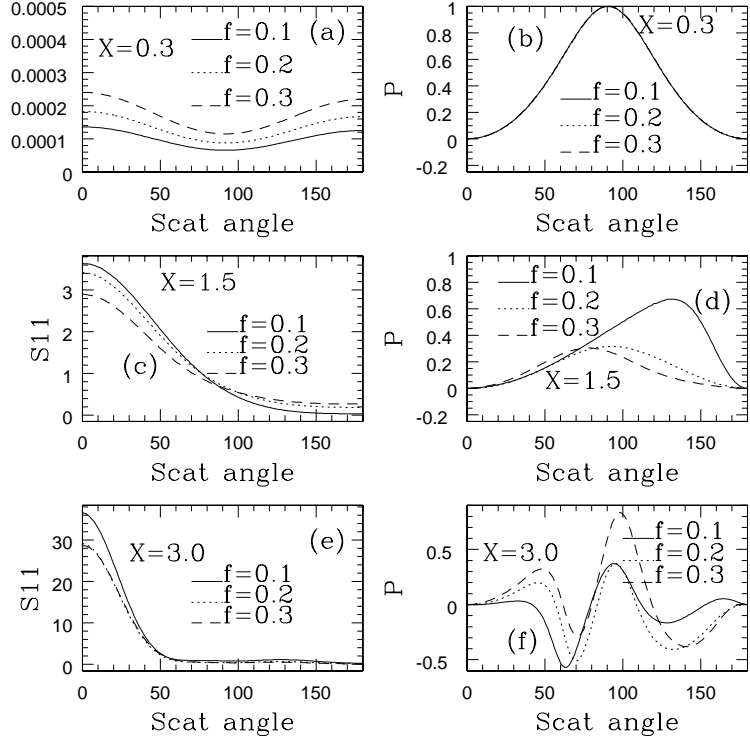}}
 \caption{Scattered intensity (S11) and Polarization P for the
  Spherical Composite grain (N=57856),
  with silicate as host sphere and graphite inclusions showing variations
  with various volume fractions.}

  \end{figure}

  It is seen that the scattered intensity S11
   for small composite grains (X $\sim$ 0.3), is increasing with
   the volume fraction of the inclusions. For
  larger grains, i.e. X $\sim$ 1.5 and 3.0 the scattered
  intensity S11 decreases with the increase in the volume fraction of the
  inclusions up to 90$^{\circ}$ and 50$^{\circ}$ scattering angles
 respectively,
   beyond these angles there does not seem to be any variation with the
   volume fractions.
  It should be noted that these large grains (X $\sim$ 1.5 and 3.0) do
not
 show any
  back scattered enhancement.
   Figure 2b shows the linear polarization for small (X $\sim$ 0.3)
grains.
  The polarization is maximum around 90$^{\circ}$ but there is no
variation
   with the volume fraction. For the grains with X $\sim$ 1.5 ,
  polarization P, shows considerable
  variation with the volume fraction of the inclusions (Figure 2d).
  It is seen that the angle of maximum polarization
  shifts from about 125$^{\circ}$ for the volume fraction f=0.1 to
 90$^{\circ}$
  for fraction f=0.2 and to 70$^{\circ}$
  for the volume fraction f=0.3 of the inclusions. The degree of
polarization
 P also decreases from about 0.6 to 0.2 as the volume fraction of the
 graphite
  inclusion increases.  For large grains,
  $\rm X=3.0$, these curves show negative polarization for all the 
volume fractions of inclusions and the positive polarization
increases with the volume fraction f of the graphite inclusions.
Figures 3(a-f) show the scattering function for spherical composite
  grain for two inclusion sizes (viz. n = 6 and 12) at a constant volume
  fraction of the inclusion (f = 0.2). It is seen that there is no
 appreciable variation in the scattered intensity S11 with the inclusion
 size.
  The scattered intensity is flat beyond 90$^{\circ}$ i.e. there is no
back
  scattering. The degree of polarization P for grains upto ( X =1.5) does
not
    show appreciable variation with the inclusion size. For larger grains
(X=3.0)
    P is higher
   for large inclusion size (i.e. n = 12) than that is obtained for n = 6.
  However, there is no shift in the position of the maximum polarization
  $\rm P_{max}$, i.e. the angle of maximum polarization, with the
inclusion
 size.

  \begin{figure}
 \centerline{\includegraphics[width=25pc]{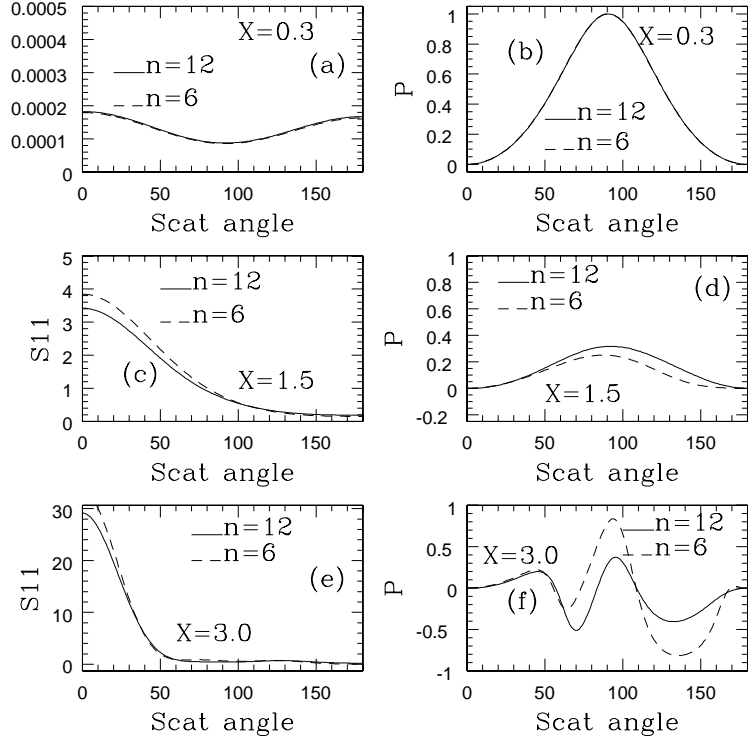}}
 \caption{Scattered intensity (S11) and Polarization P for the
  Spherical Composite grain (N=57856),
  with silicate as host sphere and graphite inclusions showing variations
  with various inclusion sizes for a fixed volume fraction of 20\%.}

  \end{figure}

We have also compared the results on the composite grains using DDA
with the results obtained using the Maxwell Garnett effective medium 
approximation (EMA) (Bohren and Huffman 1983).
Using EMA we have obtained the optical constants for the composite grains
and have used these constants in conjunction with
  Mie theory to calculate the scattered intensity and the polarization P.
 The EMA results for the composite Si+Gr
  (i.e. silicate host and graphite as inclusions) grain models
deviate considerably from DDA results (Figure 4a-f). EMA results do not
 agree with
 DDA results for Si + Gr grain models because the EMA does not take into
 account the inhomogeneities within the grain
 i.e. internal structure, voids, surface roughness etc. (see e.g. Perrin
and
 Lamy 1990, Wolff {\it {et al.}} 1994, 1998). 
 Since Maxwell-Garnet mixing rules provide the extreme cases of the
possible
 values of the effective
 dielectric constants of a two component mixture (Chylek {\it {et al.}}
2000)
 we have used the Maxwell-Garnet mixing method. Investigating the effects
of
 compositional inhomogeneities and fractal dimension on the optical
 properties of
 astrophysical dust  Bazell and Dwek (1990) have found Maxwell-Garnett
mixing
 rule
 better than the Bruggeman rule (Bohren and
Huffman 1983). However, the criteria of validity of
 respective theories
 are not clear (Bohren and Huffman 1983, Perrin and Sivan 1990).
 It would be therefore very useful and advantageous to
 compare the DDA results for the composite grains (with varying inclusion
 sizes and
 volume fractions) with those obtained by other EMA/Mie type series
solution
 techniques in
 order to examine the applicability of several mixing rules (see e.g.
Chylek 2000, Wolff {\it {et al.}} 1998).

    \begin{figure}
 \centerline{\includegraphics[width=25pc]{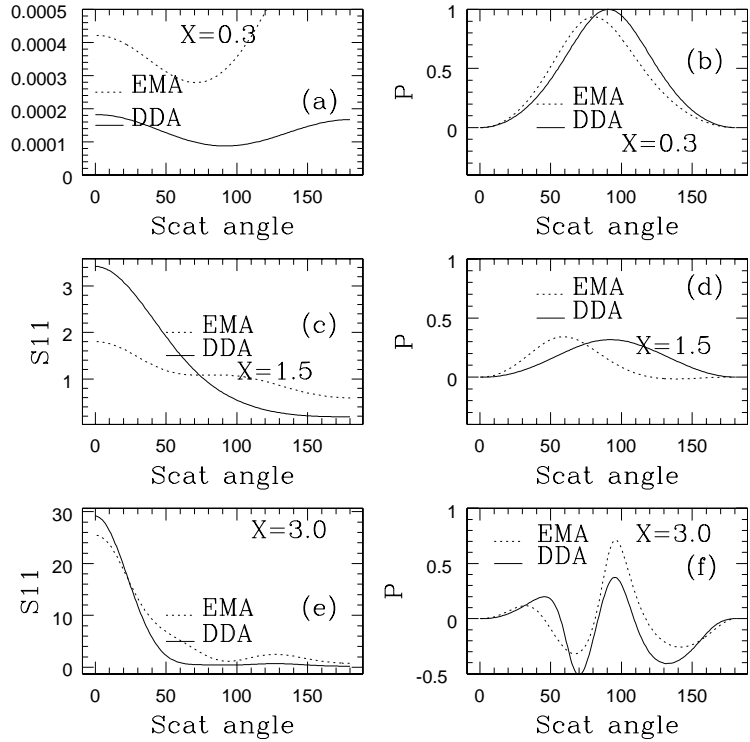}}
 \caption{Comparison of Scattered intensity (S11) and Polarization P
     for the DDA Spherical Composite grain (N=57856), Si+Gr, and EMA
      for various grain size parameters X.}

    \end{figure}

  {\it{Spheroids:}}

The main purpose of the present study is to use the results of
 the composite grains to interpret the observed scattering properties
 of the cometary dust.
 Several studies have shown that the cometary dust particles are
nonspherical
 (see for example Greenberg 1980, Jockers 1997).
 The collected interplanetary dust particles (IDPs) of likely cometary
origin
 are nonspherical (Brownlee 1978). Vaidya and Desai (1996) and Kiselev
and
 Velichko (1998) have also suggested nonspherical porous particles to
explain
 the scattering properties of cometary dust. Also, spherical particles
show
 resonances in the scattered intensity and polarization (Bohren and
Huffman
 1983) which are not present in the nonspherical particles and are not
 observed
 in cometary dust.
 Hence, in addition to the spherical grains, we have
 also studied the scattering properties for the composite spheroidal
grains.

   Figures 5 (a-f) show the scattered intensity and
   the polarization P for the composite grains containing a
   host spheroid with 14440 dipoles for three volume fractions of the
     inclusions. It is seen that for these spheroidal composite grains
  there is no appreciable variation in the scattered intensity S11 with
the
  volume fraction of the inclusions.
  The scattered intensity curves do not display any back scattering.
  The linear polarization P for small (X $\sim$ 0.3) spheroidal composite
 grains  peaks around 90$^{\circ}$ and there is no
 variation
  with the volume fractions of inclusions.
  For grains with $\rm X = 1.5$, the angle of
  maximum polarization shifts with the variation in the volume fraction
  of inclusions i.e. it shifts from 120 degree for f=0.1 to 80 degree
   for f=0.3.  Fig 5f shows linear polarization for large grains
$\rm X = 3.0$ for three volume fractions. It is seen that the composite
grain with a fraction of f=0.1 and 0.2 produce negative polarization
but with f=0.3 of graphite inclusion, the negative polarization
vanishes.

  \begin{figure}
 \centerline{\includegraphics[width=25pc]{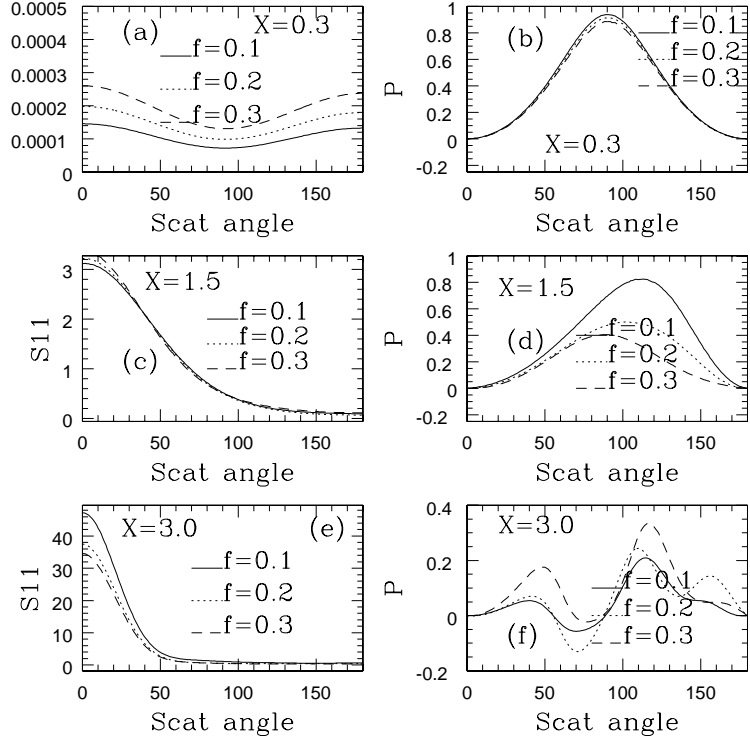}}
 \caption{Scattered intensity (S11) and Polarization P for the
  Non-Spherical Composite grain (N=14440),
  with silicate as host sphere and graphite inclusions showing variations
  with various volume fractions.}

  \end{figure}

         We have also calculated the scattered intensity for the
spheroidal
  composite grains (N=14440) for 3 inclusion sizes, viz. n = 6,8 and 12,
at
  a constant volume fraction of 20\%.
         The scattered intensity does not show variation with the
inclusion
   size (Figs. 6 a,c,e) . The polarization P for small grains , X=0.3
does
    not show any variation with the inclusion size. For the grains with X
=
1.5, the angle
  of maximum polarization $\rm P_{max}$ shifts from about $120^{\circ}$
  for n = 6 to $90^{\circ}$ for n = 12, (Fig. 6 d).
For large grains i.e. $\rm X = 3.0$ these three curves show negative
polarization for both these inclusion sizes viz. n=6 and 12 (Fig. 6f).

  \begin{figure}
 \centerline{\includegraphics[width=25pc]{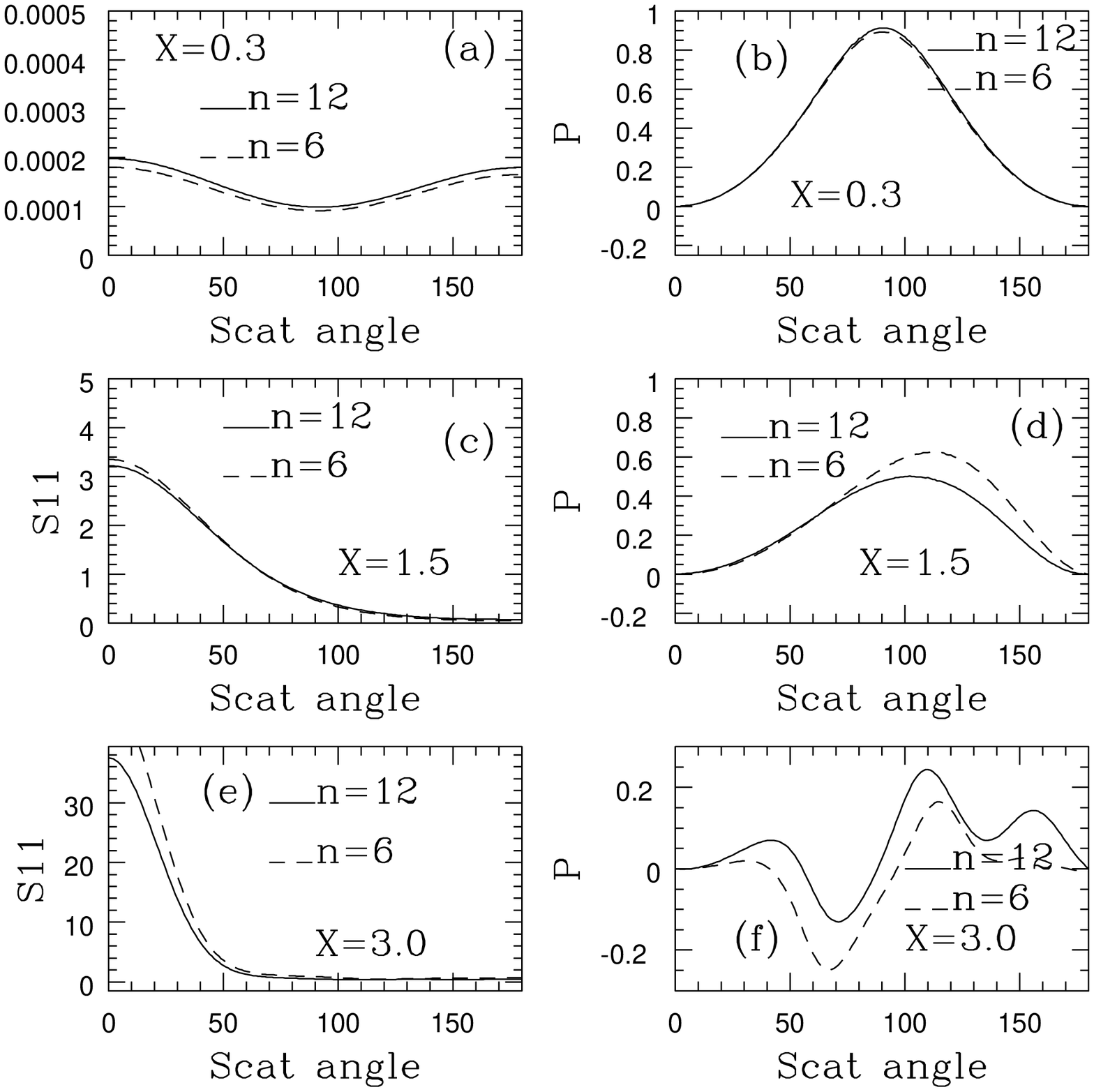}}
 \caption{Scattered intensity (S11) and Polarization P for the
  Non-Spherical Composite grain (N=14440),
  with silicate as host sphere and graphite inclusions showing variations
  with various inclusion sizes for a fixed volume fraction of 20\%.}

  \end{figure}

Figures 7a-f show 3-D plots of scattered intensity S11 and linear 
polarization P, for the composite grains with N=14440 and 152 and 
grain sizes $\rm a_{eq}=0.1-1.0\mu$ with f=0.2. 
It is seen that for the composite grains
N=14440, S11 curves are fairly flat for all the grain
sizes; however for the composite grains with N=152 and for the grain
sizes $\rm a_{eq} > 0.5 \mu$, the S11 curves show backscattering, i.e.
these curves show enhancement in the backward direction.
The linear polarization P for the composite grains sizes
 $\rm a_{eq}=0.1-0.5\mu$ 
increases with the scattering
angle, it becomes maximum around 90 degree and goes to zero at 180 degree.
The curves for larger grain sizes (i.e. $\rm a_{eq}>0.5\mu$) show
negative polarization at intermediate scattering angles before they
fold-up to P = 0 at 180 degree (fig. 7(b) \& (d)). 
It is to be noted here that the composite grains with N=152
are porous with porosity p=80\% (p is defined in section 2).
These results show that the composite
grains with high porosity can produce backscattering,which is observed
in cometary dust (Hage and Greenberg 1990b). 
Petrova {\it {et al.}} (2000) have on the other hand shown that more compact 
particles display the negative polarization and the backscattered
intensity better than as compared to porous particles.

In order to interpret the observed angular scattering and
linear polarization in cometary dust, we have used a power law size 
distribution for the composite grains.

Figures 8(a-f) show the scattered intensity S11 and linear polarization
P for the composite grains with N=14440, 1184 and 152 for a size 
distribution,
$\rm n(a)da \propto X^{-q}$, with the index q = 3.5 in the size
parameter range of X = 0.25-2.5 (i.e. $\rm a_{eq}=0.1-1.0\mu$). 
It is seen from Figure 8 (a) \& (b)
that the spheroidal grains with N=14440 and 1184 and grain size
distribution of $\rm a_{eq}=0.1-1.0\mu$ do not show back scattering. 
Figure 8(c) shows the scattered intensity for spheroidal composite 
grains with N=152, i.e. with 80\% porosity which clearly shows the 
back scattering. Linear polarization curves for N=14440 and 1184 in 
Figures 8 (d) \& (e) do not show maxima and they are fairly flat 
i.e. they are showing almost constant degree of polarization over a 
wide range of scattering angles. However, the linear polarization
curves for spheroidal grains with N=152 (Figure 8f) shows maximum 
degree of polarization at 90$^{\circ}$. 

Fig 9(a-f) show S11 and P for
smaller grain size distribution i.e. $\rm a_{eq}=0.06-0.6\mu$.
For this small size distribution also, the spheroidal composite grains
with N=14440 and 1184 do not show back scattering (see Figure 9 (a)
\& (b)), whereas the spheroidal grain with N=152 (i.e. with $\sim$
80\% porosity) does show the back scattering (Fig 9 c).
Linear polarization curves for all the three grain models show a maxima
at scattering of 90$^{\circ}$ (Fig 9 d,e and f).

Figures 10(a-d) show the 
linear polarization for the spheroidal composite grains with N=14440 
and N=25896 for a distribution of grain sizes with
 three volume fractions. 
Figure 10 (a) shows polarization for spheroidal composite 
grains (N=14440) for three volume fractions with a power law
size distribution, in the size parameter range of X = 0.17 - 1.7,
i.e. $\rm a_{eq}=0.06-0.6\mu$.
  It is seen that for all three fractions linear polarization P
increases with scattering angle, reaches maximum around 90${^\circ}$
and fold-up to zero at 180${^\circ}$. It is also seen that P decreases
with the volume fraction of the inclusions at the angle of maximum
polarization for this size distribution. In Fig 10 (b), P increases
with volume fraction f for the larger size distribution (i.e.
$\rm a_{eq}=0.1-1.0\mu$. In Fig. 10 (c) and (d) we show linear 
polarization curves for the smaller and 
larger size distribution of spheroidal grains for N=25896.

  \begin{figure}
 \centerline{\includegraphics[width=25pc]{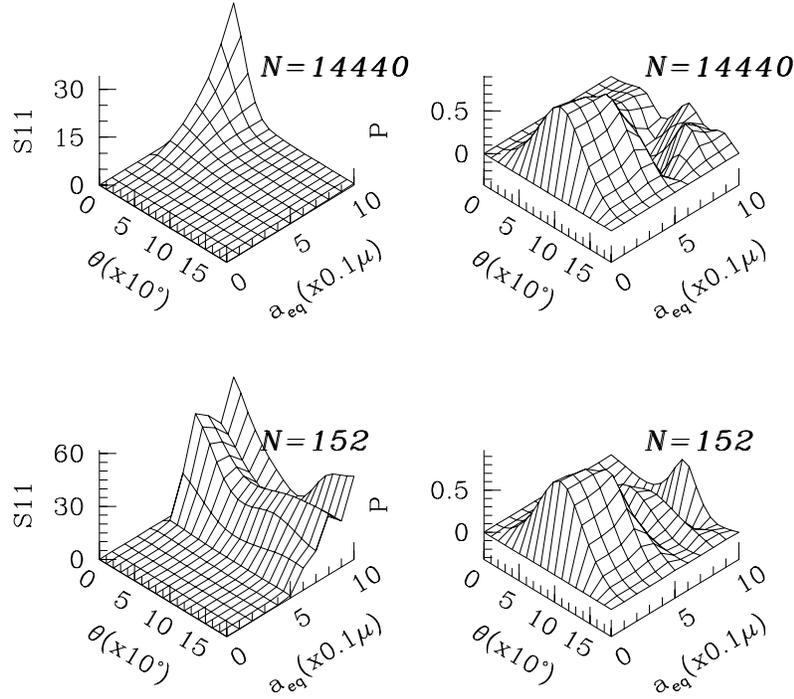}}
 \caption{3D plots of Scattering intensity S11 and
Linear Polarization (P) for Composite Spheroidal Grains with silicate
host spheroid and 20\% graphite inclusions.
The vertical axes are for
S11 and degree of Linear Polarization P; the axes marked as $\theta$ 
denotes the Scattering angle (x 10 degrees) in the range
0 to 180$^{\circ}$ and the third axis
is for grain size $\rm a_{eq}=0.1-1.0 \mu$.}
  \end{figure}

  These results on the spheroidal (Si+Gr) composite grain models with
  volume fraction of inclusions f between 0.1
  and 0.3, at infrared wavelength of $2.2{\mu m}$ show maximum positive
    polarization of about 0.3 around $90^{\circ}$ and becomes zero
at 180${^\circ}$. These results seem to be consistent
with the linear polarization observed in most of the
cometary dust (see for example Kelley {\it {et al.}} 2004).
 However, it is to be noted these composite grain models do not show
   negative branch of polarization at large scattering angles
   $\sim 140^{\circ} - 150^{\circ}$ , observed in some comets. 

 Figures 11 shows linear polarization for composite grains
with N = 25896, 14440 and 152 alongwith with the observed polarization 
curves of Comet Halley at $2.2\mu m$ 
(Brooke {\it {et al.}} 1987 represented with filled squares) and 
for observed data of comets given by Kelley {\it {et al.}} (2004) 
represented with open squares.
It should be noted that these models with a very specific
size distribution of composite grains and volume fractions
are not unique.

  \begin{figure}
 \centerline{\includegraphics[width=25pc]{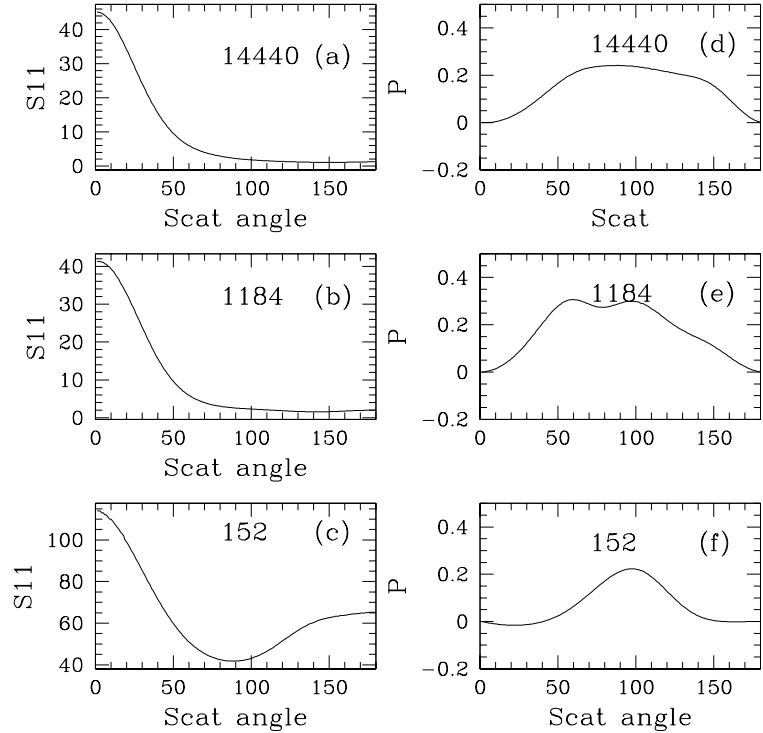}}
 \caption{Scattered intensity S11 and linear Polarization P
for the composite grain size distribution X=0.28-2.8 
($\rm a_{eq}=0.1-1.0\mu$), f=0.2.}
  \end{figure}

  \begin{figure}
 \centerline{\includegraphics[width=25pc]{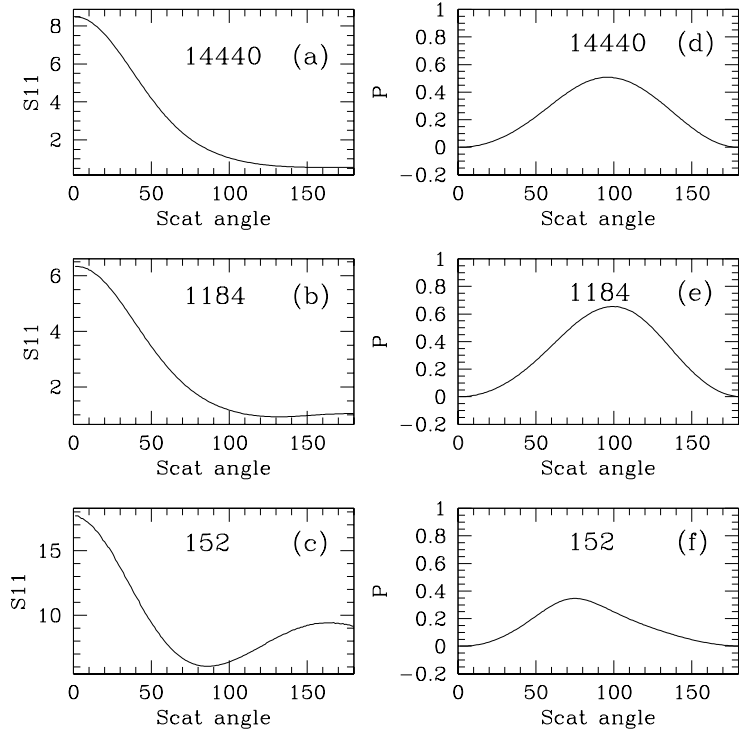}}
 \caption{Scattered intensity S11 and linear Polarization P
for the composite grain size distribution X=0.17-1.7 
($\rm a_{eq}=0.06-0.6\mu$), f=0.2.}
  \end{figure}

  \begin{figure}
 \centerline{\includegraphics[width=25pc]{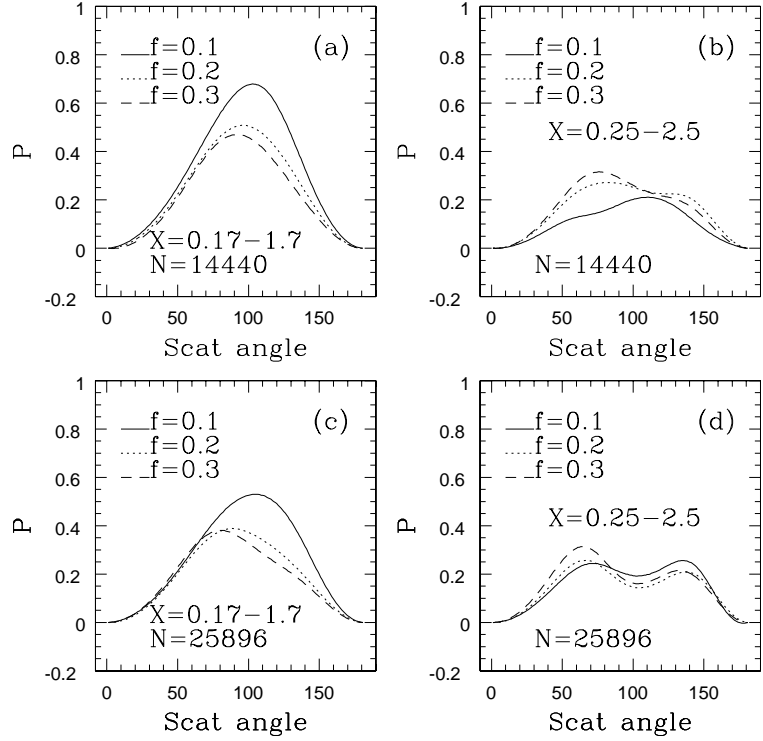}}
 \caption{Linear Polarization P for spheroidal composite grains
with size distributions $\rm a_{eq}=0.06-0.6\mu$ and 
$\rm a_{eq}=0.09-0.9\mu$.}
  \end{figure}

  \begin{figure}
 \centerline{\includegraphics[width=25pc]{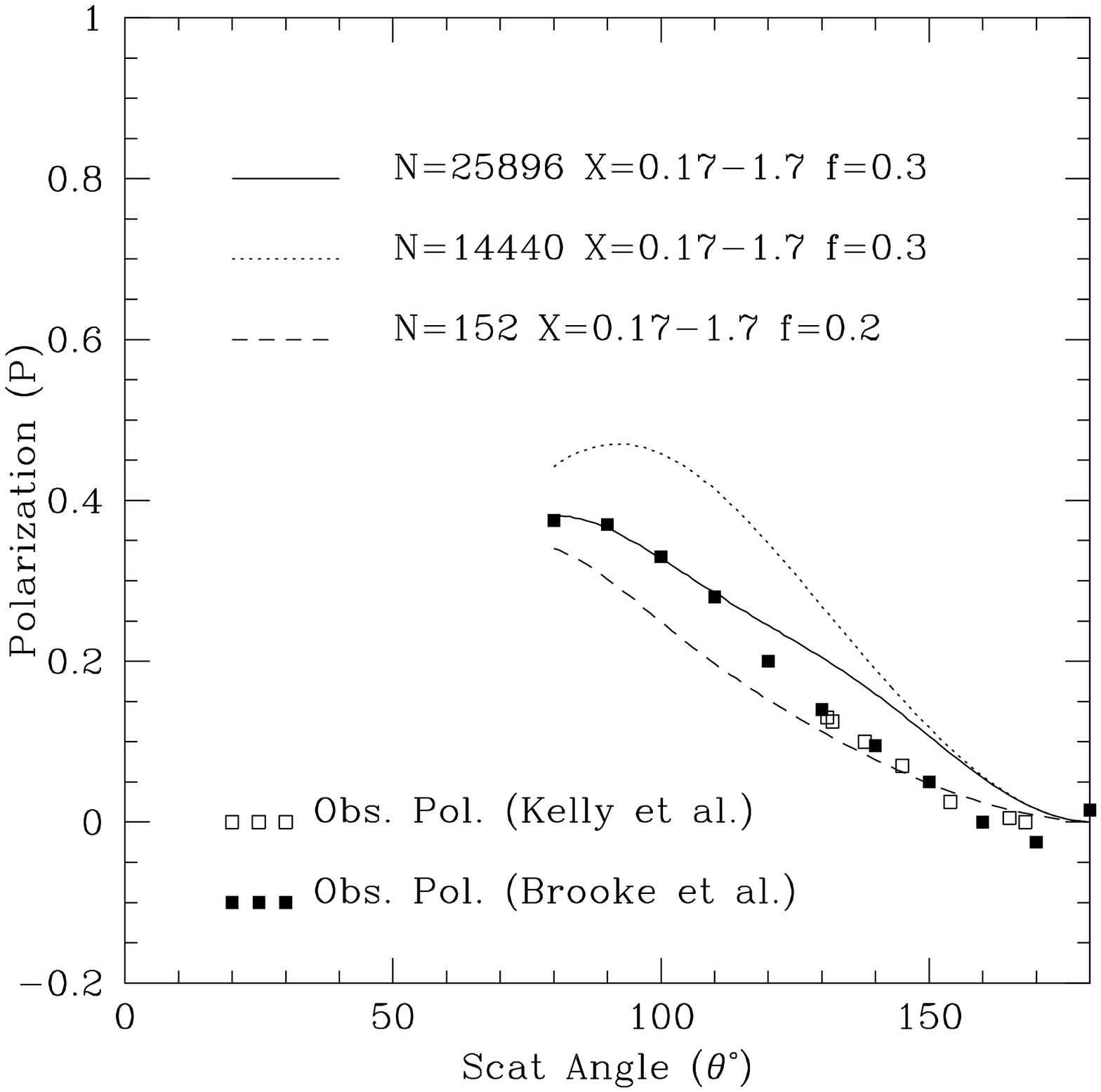}}
 \caption{Linear Polarization P for composite grain models
alongwith the observed Linear Polarization data for comets.}
  \end{figure}

 These results on the composite grains indicate that a very specific
  grain size distribution is required for a good fit to the observed
scattering and polarization curves in comets.
  Recently Petrova {\it {et al.}} (2000)
  have also studied the scattering properties of aggregate silicates and
 they
  have also found that the shape of the polarization curve is very
sensitive
 to the
  size and size distribution of the grains.
 Kerola and Larson (2001) have used T-matrix method
 to calculate the polarization upto the scattering angle 160 $^{\circ}$
 for
  prolate spheroidal crystalline olivine
 particles. Their results are compared well with the comets'
 measured polarization
 at the wavelengths 0.4845 and 0.684 $\mu m$.

 \section{Summary \& Conclusions}

 The scattering properties; i.e. the angular scattered intensity and the
  linear polarization for the composite grains with the silicate and
  graphite as the constituent materials are studied with various
inclusion
  sizes and volume fractions. We have studied these properties for
spherical
  and spheroidal composite grains.
We have also compared the results on the composite grains obtained
using DDA with those obtained using EMA.
  Our main conclusions are as follows:

  1. For small (X $\sim$ 0.3) spherical composite (Si+Gr) grains,
  the scattered intensity (S11) is found to be increasing with the volume
  fraction of the inclusions. For larger grains (X $\geq$ 1.5), S11 is
found
  to be decreasing with the volume fraction of the inclusions.
  However, the back
scattering or the scattering enhancement in the back direction
is not seen for these spherical composite grains.

  2. For small spherical grains (X $\sim$ 0.3) the linear polarization
does
not
  seem to vary with the volume fraction of the inclusion. For sizes
  X $\geq$ 1.5, $\rm P_{max}$ shifts with the volume fraction of the
  inclusions.

3. The EMA results for the composite grains (Si+Gr) deviate considerably
from the DDA results.

4. Spheroidal composite grains with silicate
   as the host material and inclusions of graphite with volume
fractions of f=0.2 and 0.3 show the maximum
positive polarization of about 0.3 around $90^{\circ}$.
        The polarization curves
  we have obtained for the spheroidal composite grains with
  axial ratios $\sim$ 1.3-2.0 in the size parameter range X=0.17-1.7, 
containing silicate as the host material
  with the inclusions of highly absorbing materials such as graphite,
  resembles the observed polarization curve in comets
  (Brooke {\it {et al.}} 1987; Kiselev and Velichko 1998;
Kelly {\it {et al.}}(2004)).
However, these composite grain models do not produce negative
polarization at large scattering angles (i.e. $>150{^\circ}$),
observed in some comets in near infra-red (NIR) wavelength region. 
It is to be noted that the so called
negative branch of polarization is not observed in
all the comets at 2.2$\mu m$ (see e.g. Kelly {\it {et al.}} 2004).

The composite grain models with very high porosity, about 80\%,
show back scattered enhancement observed in comets.
Highly porous grains with about 90\% porosity have also been suggested 
by Hage \& Greenberg (1990b) to produce the observed backscattered 
intensity.

Our results on the composite grains with silicate as the host material
and graphite inclusions show that a very specific volume fraction
of inclusion and a very specific grain size distribution may be required
to explain all the observed properties in cometary dust.
Yanamandra \& Hanner (1999) have suggested a broader grain size 
distribution to explain all the observed scattering properties
  in  cometary dust.
 It would also be interesting to study the scattering properties of the
 composite particles at few more wavelengths which would help to
 explain the polarization color effect , (i.e. higher polarization in red
 than in green) found in several comets (see e.g. Levasseur-Regourd
 {\it {et al.}} 1996).
  Laboratory scattering measurements on the composite grains at microwave
  wavelengths (e.g. Gustafson and Kolokolova 1999) would also help
  towards better interpretation of the observed scattering from cometary
  dust.

 \acknowledgements

 We thank Bruce Draine and Piotr Flatau for providing the
  DDSCAT code. DBV thanks IUCAA for continued support
  towards this research. This work on composite grains was initiated
  during DBV's visits to Dalhousie University, Halifax, Canada in years
  1999 and 2000.

 \end{article}
 \end{document}